\documentclass[12pt,preprint]{aastex}

\shorttitle{Globular Cluster/Galaxy Formation}
\shortauthors{Burstein et al.}

\begin{document}
 
\title{Globular Cluster and Galaxy Formation: M31, the Milky Way and
Implications for Globular Cluster Systems of Spiral 
Galaxies}

\author{David Burstein\altaffilmark{1},Yong Li\altaffilmark{1},
Kenneth C. Freeman\altaffilmark{2}, John E. Norris\altaffilmark{2},
Michael S. Bessell\altaffilmark{2}, Joss Bland-Hawthorn\altaffilmark{3},
Brad K. Gibson\altaffilmark{4}, 
Michael A. Beasley\altaffilmark{4}, Hyun-chul Lee\altaffilmark{4},
Beatriz Barbuy\altaffilmark{5}, John P. Huchra\altaffilmark{6},
Jean P. Brodie\altaffilmark{7}, Duncan A. Forbes\altaffilmark{4}}

\altaffiltext{1}{Department of Physics and Astronomy, Box 871504, Arizona
    State University, Tempe, AZ  85287-1504}
\altaffiltext{2}{Mt. Stromlo and Siding Spring Observatories, ANU,
Private Bag, Weston Creek, A.C.T. 2611, Australia}
\altaffiltext{3}{Anglo-Australian Observatory, P.O. Box 296, Epping,
New South Wales 2121, Australia}
\altaffiltext{4}{Centre for Astrophysics \& Supercomputing, Swinburne
University, Mail \#31, P.O. Box 218, Hawthorn, Victoria 3127, Australia}
\altaffiltext{5}{Instituto de Astronomia, Geofisica e Ci\^{e}ncias
Atomsf\'{e}ricas, Universidade de S\~{a}o Paulo, Rua de Mat\~{a}o,
1226, Cidade Universit\'{a}ria, Sao Paulo, SP, CEP 05508-090,
Brazil}
\altaffiltext{6}{Harvard-Smithsonian Center for Astrophysics, 
60 Garden Street, Cambridge, MA 02138}
\altaffiltext{7}{Lick Observatory, University of California at 
Santa Cruz, Santa Cruz, CA 95064}

\begin{abstract}

We find that the globular cluster systems of the Milky Way and of our
neighboring spiral galaxy, M31, comprise two distinct entities,
differing in three respects. First, M31 has a set of young globular
clusters (GCs), ranging in age from a few 100 Myr to 5 Gyr old, as
well as old globular clusters.  No such very young GCs are known in the
Milky Way. Second, we confirm that the oldest M31 GCs have much higher
nitrogen abundances than do Galactic GCs at equivalent metallicities.
Third, Morrison et al. have shown that M31 has a subcomponent of GCs
that follow closely the disk rotation curve of that galaxy. Such a GC
system in our own Galaxy has yet to be found.  The only plausible
scenario for the existence of the young M31 GC comes from the
hierarchical-clustering-merging (HCM) paradigm for galaxy formation.
We infer that M31 has absorbed more of its contingent of dwarf systems
in the recent past than has the Milky Way. This inference has three
implications: First, that all spiral galaxies could differ in their
globular cluster properties, depending on how many companions each
galaxy has, and when the parent galaxy absorbs them. In this spectrum
of possibilities, apparently the Milky Way ties down one end, in which
almost all of its GCs were absorbed 10-12 Gyr ago. Second, it suggests
that young globular clusters are preferentially formed in the dwarf
companions of parent galaxies, and then absorbed by the parent galaxy
during mergers.  Third, the young GCs seen in tidally interacting
galaxies might come from the dwarf companions of these galaxies,
rather than be made a-new in the tidal interaction. Yet, there is no
ready explanation for the marked difference in nitrogen abundance for
the stars in the old M31 GCs relative to those in the oldest Galactic
GCs, especially the most metal-poor GCs in both galaxies. The
predictions made by Li \& Burstein regarding the origin of nitrogen
abundance in globular clusters are consistent with what is found for
the old M31 GCs compared to that for the two 5 Gyr-old M31 GCs.
\footnote{Observations reported here were obtained at the 
MMT Observatory, a joint facility of the Smithsonian Institution 
and the University of Arizona.}

\end{abstract}


\section{Introduction}

The most studied globular system in our field of study are the
globular clusters (GCs) in our own Galaxy. These clusters have been
studied in numerous ways, including: detailed modeling of their
color-magnitude diagrams \citep[e.g.,][and references
therein]{bv01} and spectra obtained of their giant stars, horizontal
branch stars and their main sequence stars
\citep[e.g.,][]{ki03,cast00,cetal04,behr03}.  Yet, it was not until
\citet{lb03} that it was discovered that in integrated light, the most
metal-poor Galactic GCs show relatively strong molecular absorption
systems for NH at 3360$\rm \AA$.

Given all that we know about Galactic GCs, it was reasonable
to assume that the GC systems around other spiral galaxies
similar to ours would also be similar to ours: generally old GCs
having similar ages and metallicities among their constituent stars.
However, recently several groups who have studied the M31 globular 
cluster system have uncovered significant differences between that 
system and that of our own Galaxy. 

\citet{morr04} have shown that one can use accurate radial
velocities for M31 GCs to divide them into disk and bulge categories,
with the disk GCs having radial velocities that closely follow the
disk rotation curve of M31. \citet{beas04} and \citet{barm00} show 
that M31 has globular clusters that show A-type spectra, from which
one infers that these GCs are very young.  Two of us (YL and DB) have
recently shown that four of the most luminous M31 GCs have far
stronger NH features in them than do Galactic GCs measured at
otherwise similar, other metal-line indices, such as CH or Mg$_2$
\citep{lb03}.

In a follow-up to the Li \& Burstein paper, our group has
obtained spectra down to 3250$\rm \AA$ of 22 new M31 globular
clusters spanning a wider range in absolute magnitude than
the HST sample used by \citet{pon98}.  These are combined with
6 young M31 globular clusters observed by \citet{barm00} to present
a more coherent picture of the M31 GC system. Section~2 of 
the present paper details the new MMT observations. Section~3 
presents the spectra, together with spectra for the Galactic GCs
we presented in \citet{lb03}. These spectra convincingly
show that M31 contains GCs with a range of age from as young as 
$\sim$100 Myr to 5 Gyr-old, to as old as the oldest Galactic GCs.
They also clearly show the differences in nitrogen abundance between
the old M31 GCs and the old Galactic GCs.  Section~4  discusses the 
various implications of the differences found between the M31 and 
Milky Way GC systems for galaxy formation and the existence of
young globular clusters in spiral galaxies and tidally-interacting
galaxies. Section~5 contains our summary.

\section{The MMT Observations}

Li \& Burstein were assigned two nights with the 6.5m MMT on
26/27Sep03 and 27/28Sep03 to use the Blue Channel spectrograph
to obtain spectra that reach to the blue atmospheric limit of a
selection of M31 GCs.  The GCs chosen for this study were taken 
from two sources: First, it was desirous to re-observe many of the 
M31 GCs observed by \citet{bur84} in order to insure we were on the 
same absorption-line system as defined by the Lick Observatory 
data. Many of these are also among the brightest M31 GCs. For all 
but one of these Lick Observatory-observed M31 GCs, Morrison et al. 
have defined whether or not these systems follow closely the rotation 
curve of M31 or not.

Second, our remaining GC sample are taken randomly from among the
GCs that  have the ``residual'' parameter defined by Morrison et al. 
It is this ``residual'' parameter that Morrison et al. use to
measure how close or how far the radial velocity of a GC in M31
is to the rotation curve of that galaxy.  However, 
one can only be sure if a GC is a bulge GC, as there
are no disk GCs with a residual parameter greater than 2.2
\citep[e.g., as can be seen in Fig. 6 of][]{morr04}.  Hence,
while one can choose GCs that are clearly bulge GCs, choosing
disk GCs is more problematical. Given the distribution of
residuals for disk GCs in Fig. 6 of Morrison et al., we chose
those M31 GCs with residual parameters close to zero to maximize
our chances of picking up true disk GCs in M31.

We employed the 300l blue grating with the Blue Channel of the MMT
spectrograph blazed at 4800$\rm \AA$ with a $2'' \times 180''$ long
slit. (The $2''$ width to cover most of the area of these
spatially-resolved GCs, the length to provide the background for many
of these GCs, most of which was due to M31 itself.). This combination
yielded a spectral resolution close to 13$\rm \AA$.  A grating angle
of 1.12$^\circ$ was sufficient to get us to 3200$\rm \AA$ in the blue,
but still gives us out to 6000$\rm \AA$ in the red, redward of which
we run into 2nd-order confusion with the spectra.  Standard star
observations of BD+33~2642, G138-31, BD+28 4211, G191-B2B and
Feige~110 are interspersed with the program objects, to provide
calibrations for spectral energy distributions for these GCs.  Table~1
gives a log of our observations, including: the globular clusters
observed, giving their Battistini et al. (B) and Sargent et al. (S)
numbers, as well as their Vetesnik numbers (when available)
\citet{betal80,setal77,vet62}); if there are HST (H) or Burstein et
al. (L) observations of these clusters; the exposure times used for
their spectra; their apparent magnitudes (not corrected for Galactic
extinction); and their Morrison et al. residual parameters.  We adopt
a value for V mag Galactic extinction for these M31 GCs from
\citet{bur84} of $\rm A_V = 0.25$ mag.

\section{The Spectra}

As seen in Table~1 and stated above, 11 of the M31 GCs in our sample
were nominally observed by \citet{bur84}. However, as we will see when
we compare line strengths to the Lick Observatory M31 GC sample, there
is a question about the identity of one of the Lick Observatory M31
GCs. Figures 1-4 contain the spectra we obtained for these 22 M31 GCs.
Both their apparent V mag and their Morrison et al. residual parameter
is given for each GC in Figs.1-4 (when available).  Figure 5 gives the
spectra that we published in \citet{lb03} for eight Galactic globular
clusters for comparison.  Figure 6 gives the Keck spectra from
\citet{barm00} for six of the globular clusters that they indicated in
their table as being ``young?''.  Only two of the Barmby et al. GCs
shown in Figure 6 have Morrison et al. residual parameters, and both
of these are consistent with these being disk GCs.  The Keck spectra
have been ``Hanning-smoothed'' (i.e, a running average of flux of
$(0.5\times f_1 + f_2 + 0.5\times f_3)/2$). (Note that the Barmby et
al. Keck spectra do not go much below 3700$\rm \AA$.) All spectra in
Figures 1-6 are displayed in terms of $\rm \log(flux)$ versus
wavelength.  The log stretch used for all of the spectra in this paper
is 1.1 dex (versus 0.7 dex used in \citet{lb03}).

Three things are notable about these new M31 GC spectra:

1. For the oldest 12 of the new M31 GC (two are the same as previously 
observed by HST; cf. Table~1), it is obvious that the strong NH 
absorption we saw for the spectra of 4 M31 GCs in \citet{pon98} is 
also present in these new M31 GC spectra. However, there is only 
one M31 GC with a V mag fainter than 17 in this group (B001-S039), and 
its spectrum is far noisier than the other spectra in this group. 
Otherwise, all of the GCs with strong NH in their spectra have V mags 
between 13.7 and 16.5 (or, given a distance modulus to M31 of 24.38 
\citet{freed01}, absolute V mags from near -11 to -8, once Galactic 
extinction is taken into account).  Most of the 14 GCs in Figures 1 and 
2 are bulge/halo clusters, according to either their position on the 
sky (Mayall II) or their Morrison et al. residual parameter. Several
have disk-like Morrison et al. residuals, but have spectra like those
of the other bulge GCs (and, hence, are put in Figures 1-2).

2. In contrast, six of the M31 GCs we observed have Morrison et al. 
residual parameters that place them squarely within the M31 
disk system.  All six have quite remarkable spectra (Figure~3).  
All show a very strong Balmer decrement, and a strong Balmer line 
series with the broad absorption hydrogen lines typical of dwarf A
stars, and Ca II H+H$\epsilon$ much stronger than the Ca II K line.
Of these six M31 GCs, two are in common with the Barmby et
al. published sample of M31 GCs (B216-S267 and B315-S038). The
spectra of six GCs taken from the sample observed by Barmby et al. 
(Figure 6) show similar A-type stellar behavior, with a wider range 
Ca II H+K issues than those observed with the MMT. If we include all 
of the M31 GCs indicate as ``young?' by Barmby et al., this makes as 
many as 19 young GCs in M31 known so far. (However, not all of the
Barmby et al. ``young?'' GCs have Morrison et al. parameters, so it
is not clear if only disk GCs are preferentially young GCs.)

We can an estimate of how young these M31 GCs are by using the
published integrated spectra of the young GCs in the Large Magellanic
Cloud (LMC) given by \citet{lr03}. As the LMC spectra are also fluxed,
we see that the youngest LMC clusters (ages 200 Myr or younger) have
spectra that have deeper Balmer lines than the young M31 GC in
Fig. 3. Hence, an estimate of 500 Myr is not an unreasonable estimate
for the ages of these six MMT M31 GC.  It is possible that some of the
Barmby et al. clusters are somewhat younger than this age
(e.g., B380-S313 and B321-S046) and one perhaps as old as 1 Gyr
(B347-S154).  While Barmby et al. point out that others in the past
have indicated that M31 contains such young GCs, it really takes these
kind of spectra to bring this issue home to all of us.

3. The two M31 GCs in Fig. 4 (B232-S286 and B311-S033) also seem
younger than the oldest M31 GCs. Their spectra appear similar (modulo
S/N issues) with the spectra of the LMC clusters NGC 1795 and NGC 
1754 as seen in \citet{lr03}, which have estimated ages of order 5 Gyr.

We can compare the absolute magnitudes of these young M31 GCs with
their compatriots in the LMC. Using a distance modulus of 18.50 to the
LMC \citep[e.g.,][]{freed01}, the LMC GCs that have ages 200 Myr and
younger have absolute B magnitudes from $\rm M_B = -6.5$ to -9.0,
reddening-corrected. The two older, $\sim 5$ Gyr-old LMC GCs have
absolute magnitudes of -6.5 and -7.3. 

There is no evidence from the fluxed spectra shown in Figures 1-4 that
any of the M31 GCs we observed have significant reddening from inside
M31 itself, so we use the Galactic estimate of $\rm A_V = 0.25$ mag
for all of the clusters. With this value of extinction, the younger
M31 GCs in Figs. 3 and 6 have apparent V magnitudes 15.3-18.6, or
absolute mags from -9 to -5.5. This is quite similar to the range of
absolute magnitudes of the 200 Myr LMC GCs.  The absolute B magnitudes
of the two $\sim 5$ Gyr GCs are both close to -9.  From this, we think
we can conclude that the younger M31 GCs are quite analogous to the
younger LMC GCs, in that their high luminosities are not necessarily
indicative of their overall mass.  In contrast, the absolute
magnitudes of the 2, 5-Gyr-old M31 GCs indicate that they are
considerably more massive than their LMC counterparts.

\subsection{The Line Strength Measures}

Among the absorption line systems we have measured in our MMT spectra,
for the present paper we include those of NH, CH (the G-band) and
Mg$_2$.\footnote{Quantitative measurement of all of the absorption line 
systems in our spectra will be published in the Ph.D. thesis of YL.}
Measurement of the latter two indices are defined on the
Lick Observatory system \citep[e.g.,][]{bur84}, but with the caveat 
that our fluxed spectra define a different long-wavelength continuum 
than do the quartz-calibrated spectra that define the Lick system.  
We define the NH parameters according to the precepts of \citet{dc94}, 
who did use fluxed data to obtain their measurements. Further, given 
that we separately observed the Galactic globular clusters by scanning 
the telescope in a large swath over them, we also need to see if the 
transformation to the Lick system is different for the Galactic GC 
as opposed to our MMT spectra for the M31 GC.  We give both our line 
strength data and those from the Lick observations 
\citep[taken from][for consistency]{trag98} in Table~2.

We note that one of the M31 GC in Table~2 shows a marked difference in
the $\rm Mg_2$ and CH measures we obtained for it compared to what was
published in 1984: B232-S288/Vetesnik 99. Our spectra clearly show
this cluster to be very metal-poor, while the Lick spectra clearly
show it to be more metal-rich.  A check of our observing log
position for this cluster confirms that we observed the cluster we
have so-identified. Such confirmation is not available for the older Lick
data, so we assume that the M31 cluster listed as V99 in the Lick
paper is not V99. Hence, we exclude this cluster from the
line-strength comparisons with the Lick Observatory data (but not 
from other analyses).

For the eight Galactic GC, we find that the difference in magnitude for
$\rm Mg_2$, in the sense of Trager et al. minus us, is $-0.008 \pm 0.003$
average difference, with 0.008 mag error per object.  Essentially all of
the error is observational.  For the 10 remaining M31 GCs that are in
common with the Trager et al. data, we have $+0.029 \pm 0.003$ mag average
difference, or 0.010 mag error per observation. Most of the error is a 
combination of observational errors from both data sets.

In the case of CH we find an average difference for the 8 Galactic GCs
of 0.006 \AA (Trager et al. minus Us), with a mean error of 0.171 \AA, and a
single observational error of 0.484~\AA.  We note that the mean error of
the Lick observations is 0.36 \AA, more than sufficient to account for
the errors observed between these two sets of data. For the 10 M31
GCs, we get an average difference of 0.146 \AA with a mean error of
0.235 \AA with a single observational error of 0.742 \AA.  Given that
the mean difference is much smaller than the mean errors for the CH
measurements of both the Galactic GCs and the M31 GCs, we assume that
we are on the Lick Observatory line-strength system for CH for both
sets of data.

We note that to place the 2.4m Bok telescope observations of the 125 
Lick Observatory stars observed by YL for his Ph.D. thesis, offsets 
in the sense of Trager et al. minus YL are +0.006 $\pm$ 0.001 for 
Mg$_2$ and +0.211 $\pm$ 0.04 for CH.  These differences have been 
applied to the stellar data used in our figures.  The differences in 
getting to the standard Lick system for stars versus GCs is 
understandable in terms of both Mg$_2$ and CH, given the wide 
wavelength regions that define both indices, and the differences 
in spectral energy distributions between individual stars and 
integrated stellar populations. 

In Table~2 we list the absorption line values for NH, CH and Mg$_2$
for both the M31 GCs we observed here and the Galactic globular
clusters we presented in \citet{lb03}, together with their errors and
the absolute magnitudes of these clusters. In Figure~7 we present the
relationships among NH, CH and Mg$_2$ for the 17 older M31 GCs
(including the two 5-Gyr-old GCs, but not the 500-Myr-old GCs),
including the two HST-observed M31 GCs not observed by us with the
MMT, as well as for the Galactic GCs. The M31 GCs observed with the
MMT and the Galactic GCs have their Mg$_2$ corrected by adding 0.029
and -0.008 mag, respectively, to their measured values to bring them
into accord with the Lick standard system.  (We note that, given the
scale at which Figure~7 is plotted for CH, that mean offset calculated
for CH for the M31 GCs is smaller than the sizes of the symbols used
to represent these GCs.)

As one can see, the new M31 GCs that are the oldest have the same kind
of enhanced NH absorption feature as do the original 4 M31 GCs that we
investigated with HST/FOS.  It is only the two 5-Gyr-old M31 GCs that
have NH indices similar to that of Galactic GCs. Thus, we find three 
differences among the globular clusters of M31 and those of our 
own Milky Way:

1. M31 contains globular clusters with a range of age from 
$\sim 100$ Myr old to 500 Myr old to 5 Gyr old to as old as the 
oldest Galactic GCs. In other words, M31 contains GCs that are 
much younger than the known GC in our own Galaxy.

2. While both the oldest and most metal-poor M31 and Galactic GCs 
show enhanced nitrogen abundances in their integrated spectra, the
main sequence stars in the M31 GCs have markedly more nitrogen in 
their atmospheres than do the main sequence stars in either the 
Milky Way GCs or most field stars in the Milky Way that are of similar 
CH or Mg$_2$ line strengths.

3. As \cite{morr04} have shown, M31 has a set of globular clusters
that closely follow the rotation curve of M31.  None of the known
GCs in the Milky Way do this.

\subsection{Keeping Score with the Predictions}

One of the predictions of \cite{lb03} from their GC and halo formation
scenario is that younger GCs in other galaxies should have NH in 
proportion to their normal metallicity, as compared to the enhanced
NH we see in the oldest clusters. In this regard, we note that the
two 5 Gyr-old GCs in M31 do have NH much weaker than their 10-Gyr-old
GC cousins, consistent with this prediction. Note, however, that even
though the 5 Gyr-old M31 GCs are weaker in NH than their older
cousins, they are of comparable NH strength relative to the NH measures
for the old Galactic GCs. This is perhaps an indication that nitrogen 
is greatly enhanced in M31 GCs in general, not just in their oldest 
GCs. This makes one wonder whether older field stars in M31 have 
enhanced nitrogen abundances relative to Galactic field stars.

Separately, as pointed out by \citet{tl84}, CH (the G-band) and NH
have very similar dissociation energies. The fact that CH lies
1000$\rm \AA$ to the red of NH implies that the weakness of CH in the
integrated spectra of the M31 GCs relative means that carbon is very
under abundant relative to nitrogen in these GCs (and also, in the
Galactic GCs).  Such was also found for main sequence stars in several
Galactic GCs \citep[e.g.][]{bcp03}.  If these abundance differences
are truly primordial in origin, are not these abundance anomalies
telling us something about the elements the first stars produced?

\section{The Hierarchical-Clustering-Merging Paradigm and its
Implication for Globular Clusters of Other Galaxies}

The hierarchical-clustering-merging (HCM) paradigm is the current
view of galaxy formation \citep[e.g.,][]{bur97,kauf97}.  HCM 
pictures galaxy formation as small galaxies combining to form
larger galaxies. In an HCM-dominated universe, each galaxy will 
undergo a series of major and minor mergers in its lifetime.  If 
the merger process does not produce an E or S0 galaxy, then it will 
produce a spiral galaxy. Since there are more spirals than 
E/S0 galaxies in the Universe (e.g., the galaxies in both the Uppsala 
General Catalog, \citet{nil73}, and in the Third Reference Catalog of 
Bright Galaxies, \citet{dev91}, are dominated by
spirals), it is clear that if HCM is the dominant mode of galaxy 
formation, spirals are its preferred output. In sum, the current
evidence is consistent with the hypothesis that the HCM paradigm
applies to all giant galaxies, Es, S0s and spirals \citep[e.g.,][]{bur97}.

This, then begs a series of questions that follow in logical steps:
1. How does the HCM paradigm relate to young M31 GCs that we find in M31
but do not find in our own Galaxy?  2. Where else do we find young
globular clusters in the Local Group? 3. Where do we find most of the 
dwarf systems in the Local Group? 4. Why do young GCs preferentially 
form in irregular or small spiral galaxies? 5. Where do the young
GCs seen in tidally interacting galaxies come from?

The second question is the most obvious to answer.  We see young GCs
in the irregular companion galaxies to the Milky Way, the Large and
Small Magellanic Clouds, as well as likely in the low luminosity Sc
galaxy in the Local Group, M33 \citep[e.g.][]{ma01}.  Indeed, we have
used the integrated SEDs of LMC young GCs obtained by \citet{lr03} to
estimate the ages of our younger M31 GCs.

The answer to the third question comes from the fact that 98\% of 
the mass and luminosity of galaxies in the Local Group is contained 
in just M31 and the Milky Way.  If you plot the known positions of 
the dwarf galaxies in the Local Group (data kindly given DB by Eva Grebel),
one sees that both M31 and the Milky Way have close contingents of
these dwarf galaxies (e.g., the LMC, SMC, and various dEs and dSphs
close to the Milky Way).

Put the answers together to the previous two questions and we have our
answer to the first question. The young GCs seen in M31 likely came
from the small spiral and/or irregular galaxies that once were
companions of M31 and have since merged with M31.  Since the LMC is
less than 1\% the mass of M31, mergers with LMC-like irregulars would
do little to the disk of this galaxy. That such a merger
history has taken place in M31 is consistent with what \citet{brow03}
have found in their deep HST investigation of the halo of M31. Namely,
they find an intermediate-aged stellar population, which could be the
remnants of the dwarf system that produced the 5 Gyr-old GCs in
M31. Furthermore, the fact that one sees what has been interpreted as
a severe warp in front on the bulge of M31 \citep{bs82,bb84} now can
also be interpreted as the wrap-around debris from merger remnants.

If a galaxy similar to the LMC were accreted by M31 during the past
100-200 Myr, then all of the GCs in that irregular galaxy would 
now become GCs of M31, would they not?  And, if that accretion
took place such that the irregular galaxy would be tidally disrupted
along the disk of M31, most of the GCs in this irregular galaxy
would assume the rotation velocity of M31, forming its thin disk
of GCs. And, it is in this thin disk of GCs that we find the 
young GCs of M31. (However, as noted above, these may not be 
the only types of disk GCs in M31.)

If this is true for M31, why is it not true for the Milky Way?  
Because, in the HCM paradigm, what the merger history is of one 
galaxy does not predict what the merger history of another galaxy 
will be, even if they are close neighbors!  Evidently, the Milky
Way has not had a merger of an irregular galaxy like the LMC or
the SMC in the past 10 billion years or so, else we would see younger
GCs in our own Galaxy.

Is it not likely that many, if not most, of the GCs in spiral
galaxies were once in the smaller systems that combined to make
the spirals we see today?  If this is true, it is also true that
the HCM paradigm does not dictate how and when such accretion will 
take place. As such, we feel it is a prediction of the HCM
paradigm that the GC systems of all spiral galaxies (and for 
that matter, also gE and S0 galaxies) are assembled in a rather
haphazard ``big-fish-eats-small-fish'' manner. We see this happening
today in our own Galaxy with the Sagittarius dwarf galaxy 
\citep[e.g.,][]{ibat94} and the Canis Major dwarf \citep{fsb04}, both
of which we are accreting, each adding 4-6 GCs to our GC contingent.
(We note that two of GCs the Milky Way is accreting from each 
dwarf galaxy have ages $\sim 7$ Gyr; e.g., \citet{fsb04}.)

Why do small spirals and irregulars preferentially form young GCs?
We think this is because in such galaxies one does not have 
substantial rotational shear.  This permits large molecular clouds 
\citep[e.g., with masses up to $10^9$ $M_\odot$][]{hp94}, to form 
that will not be sheared into smaller systems. In the Milky Way such 
sheared-stellar systems tend to form the open clusters we find in our 
disk.  If this interpretation is correct, then the fact that M31 has 
both relatively young GCs ($\sim 500$ Myr) as well as moderately 
old GCs ($\sim 5$ Gyr) strongly suggests that M31 went at least a 
series of separate merging events with its contingent of small 
companion galaxies that contained many more GCs than the two dwarf
galaxies that the Milky Way is currently absorbing (e.g, the 
Sagittarius dwarf galaxy and the Canis Major dwarf; \citet{ibat94} 
and \citet{fsb04}).

If our interpretation of the origin of the young GCs in M31 is
correct, then it shows that we might be able to study the relatively
recent (say, the last 10 Gyr or so) merger history of other spiral 
galaxies by measuring the ages of their globular clusters. This 
possibility needs to be tested.

If young GCs are preferentially formed in small galaxies with little
or no net rotation velocities, then where do the young GCs that have
been found around the Antennae galaxies \citep[e.g.,][]{ws95} come
from? As an alternative hypothesis to these young clusters having been
formed during the tidal interaction, what is the chance that either of
the Antennae galaxies (NGC 4038/9) had a one or two irregular galaxies
containing young GCs that were ``brought along for the ride''?  Until
we know more about where and how young globular clusters are formed
outside of the Local Group, this idea is consistent with what we see
for tidally interacting galaxies.

The difference between M31 and Milky Way GCs that we cannot explain 
is why the GCs in M31 have markedly stronger NH absorption than do 
Galactic GCs of similar metalliticies. To-date, the M31 GCs for which 
we have found very strong NH absorptions have luminosities ranging 
from -11 (MII) to -8.5 to -9 (with apparent V mags of 15.5-16.0).  
Ironically, all of the less luminous M31 GCs we took with the MMT 
are of the young kind.

However, even if this an issue of the luminosities of the oldest 
GCs in spiral galaxies, this still begs the question of how one
gets a wide range of metallicity among these luminous GCs that spans
a similar range as among the Galactic GCs.  Of the three known 
differences among the GC systems of M31 and the Galaxy, it is the 
difference in nitrogen abundance in them that is still the most 
puzzling and opens up a number of questions that need answers.

\section{Summary}

In this paper we present evidence that the globular cluster
systems of the Milky Way and of M31, the two large spirals in the
Local Group, have very different evolutionary histories. Whereas
the Milky Way globular clusters are uniformly very old (10-12 Gyr),
those of M31 evidence at least three separate age epochs: 
100-500 Myr, 5 Gyr and 10-12 Gyr old.  We strongly suspect that 
the younger GCs in M31 came from mergers of M31 with its associated 
irregular galaxies in the past. We note that it might be happenstance 
that the Milky Way has not yet absorbed the Magellanic Clouds, for if 
it did, our Galaxy would also have a large number of young globular 
clusters in it.

We now find that there are at least three clear differences between
the GCs in M31 and those in the Milky Way: 

1. M31 contains globular clusters with a wide range in age, the
Milky Way does not.

2. While both Milky Way and M31 GCs both show enhanced nitrogen
abundances, the nitrogen abundance of the M31 GCs is clearly
greatly enhanced relative to that seen in the Milky Way GCs.
Why this is the case is still unknown. Is it possible that
nitrogen is overabundant in M31 as a system, as opposed to just
in its old GCs?

3. Morrison et al. find that M31 contains a subset of GCs whose radial
velocities closely follow the disk rotation curve of M31.  All of the
500 Myr-old GCs we and Barmby et al. find in M31 that have Morrison et
al. residual parameters (about half) are part of these disk clusters. 
No such GC component of the Milky Way has yet been discovered.

From this evidence, it is clear that if galaxies are formed via the
hierarchical-clustering-merging (HCM) paradigm, then it is likely the
case that each spiral galaxy has its own contingent of globular
clusters acquired from its contingent of dwarf systems which, in
principle, can span a wide range in age. Furthermore, it also pegs
the Milky Way at one end of this spectrum, in that almost all of
the GCs in the Milky Way are old, implying that the vast majority of
such mergers occurred more than 10 Gyr for our Galaxy.  As such, we 
have a very biased view of GC formation history in spiral galaxies 
through the study of the GCs in our own Galaxy. It will only be by 
more thorough spectroscopic investigation of the GCs in spiral (and, 
also elliptical?) galaxies outside the Local Group that the full 
extent of what the HCM paradigm means for the GC systems of galaxies 
will be understood.

\acknowledgements

DB and YL would like to thank the telescope operators at the 
MMT for their help, and the anonymous referee for helpful comments.  
DB thanks Eva Grebel for sending him her data on distances of 
Local Group galaxies.

\clearpage

\begin{deluxetable}{lrlrlr}
\tablecaption{Basic Information for 22 M31 Globular Clusters}
\tablehead{
\colhead{Globular Cluster\tablenotemark{a}} & 
\colhead{V Mag}\tablenotemark{b} & 
\colhead{Lick/HST\tablenotemark{c}} & 
\colhead{Date Obs} & 
\colhead{Exp Time} & 
\colhead{Resid Param\tablenotemark{d}}
}
\startdata
B000-S001 = May II = G1 & 13.7 & H,L      & 27/28Sep03 & 600s  & Bulge \\
B001-S039               & 17.1 & $\ldots$ & 27/28Sep03 & 1800s & 5.67  \\
B019-S072 = V44         & 14.6 & $\ldots$ & 27/28Sep03 & 1200s & 4.07  \\
B029-S090 = V29         & 16.3 & L        & 27/28Sep03 & 2700s & -0.32 \\
B158-S213 = V64         & 14.5 & L        & 27/28Sep03 & 1200s & 1.94  \\
B171-S222 = V87         & 15.0 & L        & 26/27Sep03 & 1800s & -0.04 \\
B179-S230 = V92         & 15.2 & L        & 27/28Sep03 & 1200s & 0.72  \\
B193-S244 = V116        & 15.3 & L        & 27/28Sep03 & 1200s & 0.65  \\
B210                    & 16.8 & $\ldots$ & 26/27Sep03 & 3600s & 0.11  \\
B216-S267 = V119        & 17.6 & $\ldots$ & 26/27Sep03 & 3600s & -0.10 \\
B218-S272 = V101        & 14,7 & L        & 27/28Sep03 & 1200S & 0.36  \\    
B223-S278               & 15.3 & $\ldots$ & 26/27Sep03 & 1800s & -0.01 \\
B225-S280 = V282        & 14.3 & H,L      & 27/28Sep03 & 1200s & 1.43  \\
B232-S286 = V99         & 15.5 & L        & 27/28Sep03 & 1800s & 2.52  \\
B238-S301 = V108        & 16.5 & $\ldots$ & 27/28Sep03 & 2700s & 5.80  \\
B311-S033 = V4          & 15.5 & L        & 26/27Sep03 & 1800s & 2.15  \\
B315-S038 = V5          & 16.3 & $\ldots$ & 26/27Sep03 & 3600s & -0.18 \\
B338-S076 = V12         & 14.4 & L        & 27/28Sep03 & 1200s & 4.85  \\
B386-S322               & 15.6 & $\ldots$ & 26/27Sep03 & 2700s & -7.57 \\
B400-S343               & 16.4 & $\ldots$ & 26/27Sep03 & 3600s & -8.47 \\
B484-S310               & 18.6 & $\ldots$ & 27/28Sep03 & 4500s & 0.59  \\
V31                     & 17.0 & $\ldots$ & 26/27Sep03 & 3600s & 0.02  \\
\enddata
\tablenotetext{a}{The first designations conform to the Barmby et al.
M31 GC designations; May II = Mayall II, which is also called G1 in some
papers; V = Vetesnik number.}
\tablenotetext{b}{The observed V magnitude of the cluster, taken either
from the Sargent et al. list (as given in Burstein et al. 1984), 
or the Barmby et al. list.}
\tablenotetext{c}{L = also observed by Burstein et al. 1984; H = also
observed by Ponder et al. 1998.} 
\tablenotetext{d}{The residual parameter of Morrison et al.; a value
less than $\pm 2.2$ makes it possible that this is a disk GC in M31.
However, not all M31 GCs with such residual parameters are disk GCs,
as is evident in our own data.}
\end{deluxetable}

\clearpage

\begin{deluxetable}{lrrrrrrrrrrrrr}
\rotate
\tabletypesize{\tiny}
\tablecaption{The Data for 22 M31 and 8 Galactic Globular Clusters}
\tablehead{
\colhead{Globular Cluster} &
\colhead{NH, US} &
\colhead{e(NH)}&
\colhead{CH, US} &
\colhead{e(CH)} &
\colhead{Mg$_2$, US} &
\colhead{e(Mg2)} &
\colhead{CH, Lick} &
\colhead{e(CH)} &
\colhead{Mg$_2$, Lick} &
\colhead{e(Mg2)} &
\colhead{$\rm \Delta Mg_2$} &
\colhead{$\rm \Delta CH$} &
\colhead{$\rm M_V$}
}
\startdata
b000-s001 & 7.381 & 0.218 & 3.723 &  0.065 & 0.096 & 0.001 & 3.33 & 0.34 & 0.142 & 0.008 & 0.046 & -0.393 & -10.8 \\
b001-s039 & 3.548 & 2.601 & 5.225 &  0.354 & 0.102 & 0.006 &$\ldots$&$\ldots$&$\ldots$&$\ldots$&$\ldots$&$\ldots$& -7.4 \\
b019-s072 & 7.469 & 0.322 & 4.202 &  0.082 & 0.102 & 0.002 &$\ldots$&$\ldots$&$\ldots$&$\ldots$&$\ldots$&$\ldots$& -9.9 \\
b029-s090 & 6.642 & 1.355 & 5.823 &  0.240 & 0.177 & 0.005 & 7.35 & 0.59 & 0.212 & 0.013 & 0.035 &  1.527 &  -8.2 \\ 
b158-s213 & 7.647 & 0.237 & 4.177 &  0.065 & 0.098 & 0.001 & 4.14 & 0.41 & 0.132 & 0.009 & 0.034 & -0.037 & -10.0 \\
b171-s222 & 7.057 & 0.490 & 4.976 &  0.110 & 0.175 & 0.002 & 4.53 & 0.48 & 0.189 & 0.011 & 0.014 & -0.446 &  -9.5 \\
b179-s230 & 6.066 & 0.478 & 3.938 &  0.116 & 0.071 & 0.002 & 4.39 & 0.46 & 0.104 & 0.010 & 0.033 &  0.452 &  -9.3 \\
b193-s244 & 7.858 & 0.521 & 4.810 &  0.110 & 0.197 & 0.002 & 4.07 & 0.48 & 0.211 & 0.011 & 0.014 & -0.740 &  -9.2 \\
b218-s272 & 7.119 & 0.255 & 4.086 &  0.072 & 0.096 & 0.002 & 3.78 & 0.40 & 0.123 & 0.009 & 0.027 & -0.306 &  -9.8 \\
b225-s280 & 8.550 & 0.196 & 4.639 &  0.053 & 0.146 & 0.001 & 4.88 & 0.30 & 0.184 & 0.006 & 0.038 &  0.241 & -10.2 \\
b238-s301 & 6.349 & 0.578 & 4.378 &  0.142 & 0.095 & 0.003 &$\ldots$&$\ldots$&$\ldots$&$\ldots$&$\ldots$&$\ldots$& -8.0 \\
b338-s076 & 6.203 & 0.208 & 3.126 &  0.062 & 0.053 & 0.001 & 3.03 & 0.39 & 0.085 & 0.009 & 0.032 & -0.096 & -10.1 \\ 
b386-s322 & 6.472 & 0.349 & 3.659 &  0.093 & 0.060 & 0.002 &$\ldots$&$\ldots$&$\ldots$&$\ldots$&$\ldots$&$\ldots$& -8.9 \\ 
b400-s343 & 5.570 & 0.445 & 3.672 &  0.116 & 0.055 & 0.002 &$\ldots$&$\ldots$&$\ldots$&$\ldots$&$\ldots$&$\ldots$& -8.1 \\
\tableline \\
b232-s286 & 3.399 & 0.354 & 1.891 &  0.096 & 0.003 & 0.002 & 1.55 & 0.48 & 0.073 & 0.011 & 0.070 & $\ldots$ &  -9.0 \\
b311-s033 & 3.974 & 0.447 & 1.856 &  0.113 & 0.002 & 0.002 & 3.11 & 0.53 & 0.022 & 0.011 & 0.020 &  1.254   &  -9.0 \\
\tableline \\
b484-s310 &$\ldots$&$\ldots$&$\ldots$&$\ldots$&$\ldots$&$\ldots$&$\ldots$&$\ldots$&$\ldots$&$\ldots$&$\ldots$&$\ldots$&  -5.9 \\
b216-s267 &$\ldots$&$\ldots$&$\ldots$&$\ldots$&$\ldots$&$\ldots$&$\ldots$&$\ldots$&$\ldots$&$\ldots$&$\ldots$&$\ldots$&  -6.9 \\
v031      &$\ldots$&$\ldots$&$\ldots$&$\ldots$&$\ldots$&$\ldots$&$\ldots$&$\ldots$&$\ldots$&$\ldots$&$\ldots$&$\ldots$&  -7.5 \\
b210      &$\ldots$&$\ldots$&$\ldots$&$\ldots$&$\ldots$&$\ldots$&$\ldots$&$\ldots$&$\ldots$&$\ldots$&$\ldots$&$\ldots$&  -7.7 \\
b223-s278 &$\ldots$&$\ldots$&$\ldots$&$\ldots$&$\ldots$&$\ldots$&$\ldots$&$\ldots$&$\ldots$&$\ldots$&$\ldots$&$\ldots$&  -9.2 \\
b315-s038 &$\ldots$&$\ldots$&$\ldots$&$\ldots$&$\ldots$&$\ldots$&$\ldots$&$\ldots$&$\ldots$&$\ldots$&$\ldots$&$\ldots$&  -8.2\\

\enddata
\end{deluxetable}

\begin{deluxetable}{lrrrrrrrrrrrrr}
\rotate
\tabletypesize{\tiny}
\tablecaption{The Data for 22 M31 and 8 Galactic Globular Clusters (continued)}
\tablehead{
\colhead{Globular Cluster\tablenotemark{a}} &
\colhead{NH, US} &
\colhead{e(NH)\tablenotemark{b}}&
\colhead{CH, US\tablenotemark{c}} &
\colhead{e(CH)} &
\colhead{Mg$_2$, US\tablenotemark{d}} &
\colhead{e(Mg2)} &
\colhead{CH, Lick\tablenotemark{e}} &
\colhead{e(CH)} &
\colhead{Mg$_2$, Lick} &
\colhead{e(Mg2)} &
\colhead{$\rm \Delta Mg_2$} &
\colhead{$\rm \Delta CH$} &
\colhead{$\rm M_V$\tablenotemark{f}}
}
\startdata
M53 = NGC 5024 & 3.262 & 0 & 1.328 &  0     & 0.048 & 0     & 1.32 & 0.41 & 0.039 & 0.010 & -0.009 & -0.008 & -8.7 \\
M3  = NGC 5272 & 3.476 & 0 & 1.912 &  0     & 0.061 & 0     & 2.44 & 0.45 & 0.040 & 0.008 & -0.021 &  0.528 & -8.9 \\
M5  = NGC 5904 & 4.022 & 0 & 2.351 &  0     & 0.067 & 0     & 2.31 & 0.44 & 0.067 & 0.010 &  0.000 & -0.041 & -8.8 \\
M13 = NGC 6205 & 3.766 & 0 & 1.268 &  0     & 0.055 & 0     & 1.82 & 0.20 & 0.039 & 0.005 & -0.016 &  0.552 & -8.7 \\
M92 = NGC 6341 & 2.612 & 0 & 0.618 &  0     & 0.032 & 0     & 0.87 & 0.23 & 0.021 & 0.006 & -0.011 &  0.252 & -8.2 \\
M71 = NGC 6838 & 4.767 & 0 & 5.127 &  0     & 0.159 & 0     & 4.17 & 0.45 & 0.157 & 0.010 & -0.002 & -0.957 & -5.6 \\
M15 = NGC 7078 & 2.621 & 0 & 0.626 &  0     & 0.031 & 0     & 0.63 & 0.26 & 0.023 & 0.007 & -0.008 &  0.004 & -9.2 \\
M2  = NGC 7089 & 4.099 & 0 & 1.966 &  0     & 0.053 & 0     & 1.69 & 0.44 & 0.053 & 0.008 &  0.000 & -0.276 & -9.0 
\enddata

\tablenotetext{a}{The GCs are listed in the following way: the first 14 are the old, NH-rich M31 GCs; the next
two are the 5-Gyr-old GCs; the next six are the 6 500-Myr-old GCs, and the final 8 are the Galactic GCs. No
absorption line strengths are given for the 500-Myr-old M31 GCs. M31 GC names as in Table~1. Galactic GC names given 
as both Messier and NGC numbers.}
\tablenotetext{b}{NH equivalent widths are measured in angstroms.}
\tablenotetext{c}{CH equivalent widths are measured in angstroms.}
\tablenotetext{d}{Mg$_2$ values are measured in magnitudes.}
\tablenotetext{e}{No Lick data is given if not observed by Burstein et al. 1984.}
\tablenotetext{f}{Absolute V magnitude, extinction corrected using the extinction values given in Harris 2003,
and a value of E(B-V) = 0.08 for M31 GCs.}

\end{deluxetable}

\clearpage

\begin{figure}
\epsscale{0.8}
\plotone{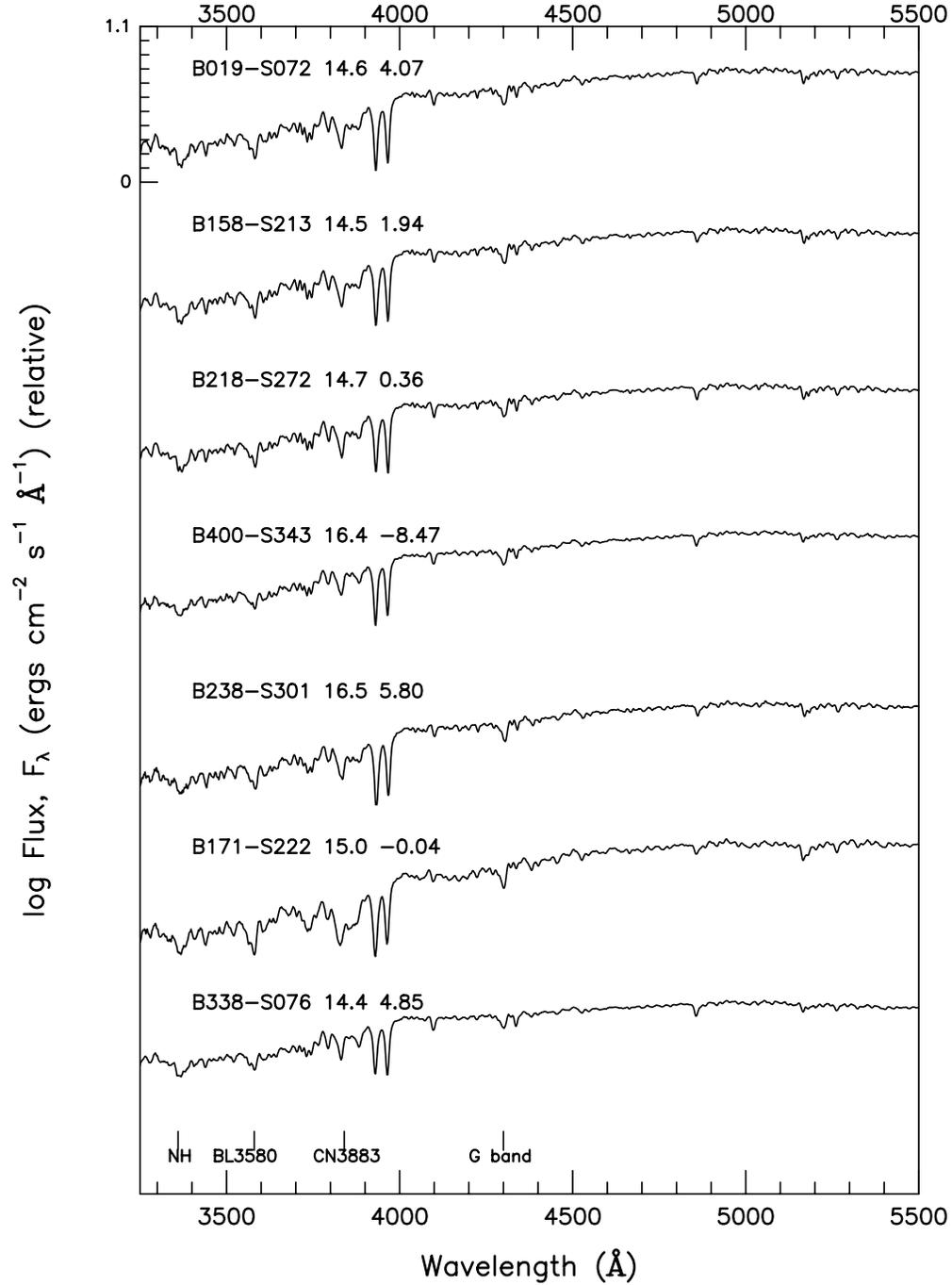}
\figcaption{Spectra for seven of the 14 M31 GCs observed with the MMT
that are both as old as the oldest Galactic GCs and have strong NH in
their spectra. Their V magnitudes and Morrison et al. residual
parameters are given by their names. The fluxes for these spectra 
are plotted on a 1.1 dex log scale }
\label{fig1}
\end{figure}

\begin{figure}
\epsscale{0.8}
\plotone{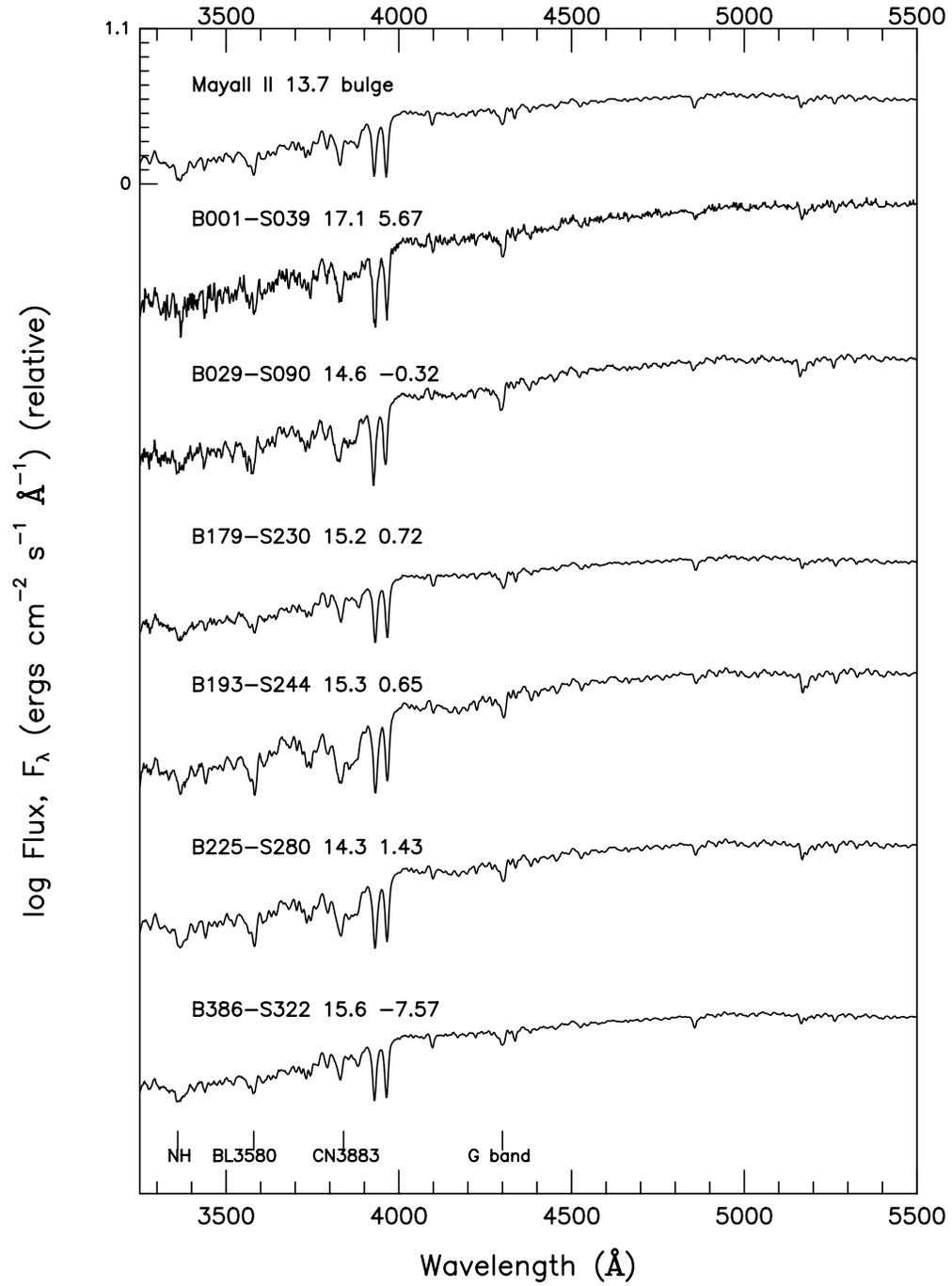}
\figcaption{Spectra for the seven other old M31 GCs, plotted on the
same log flux scale as in Figure~1.}
\label{fig2}
\end{figure}

\begin{figure}
\epsscale{0.8}
\plotone{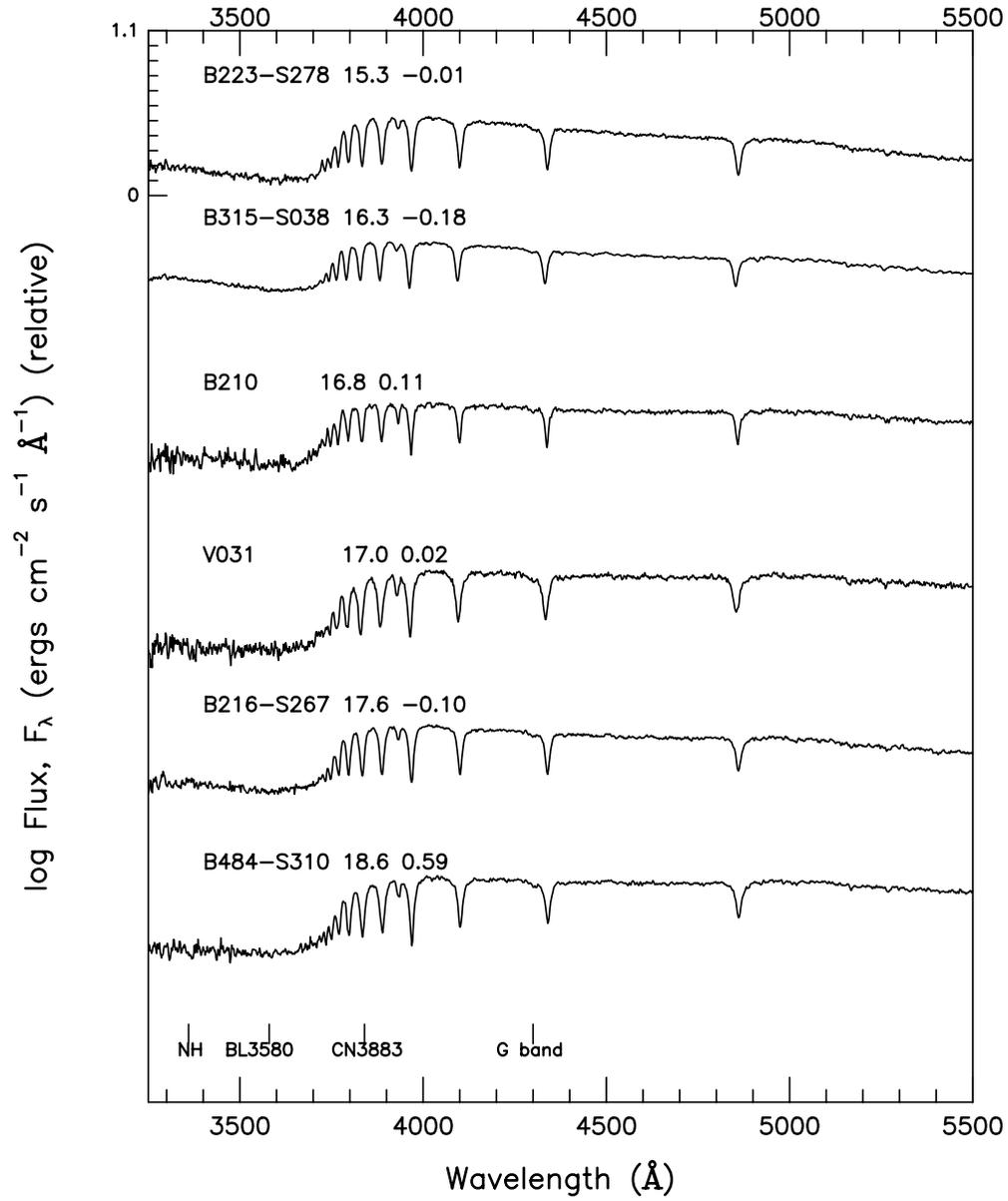}
\figcaption{The spectra for the six 500-Myr-old M31 GCs, plotted on
the same log flux scale as in Figure~1. Note the strong Balmer jump
for these GCs, as well as the strong Balmer line series in their
spectra. The log flux scale used is the same as in Figure~1.}
\label{fig3}
\end{figure}

\begin{figure}
\epsscale{0.8}
\plotone{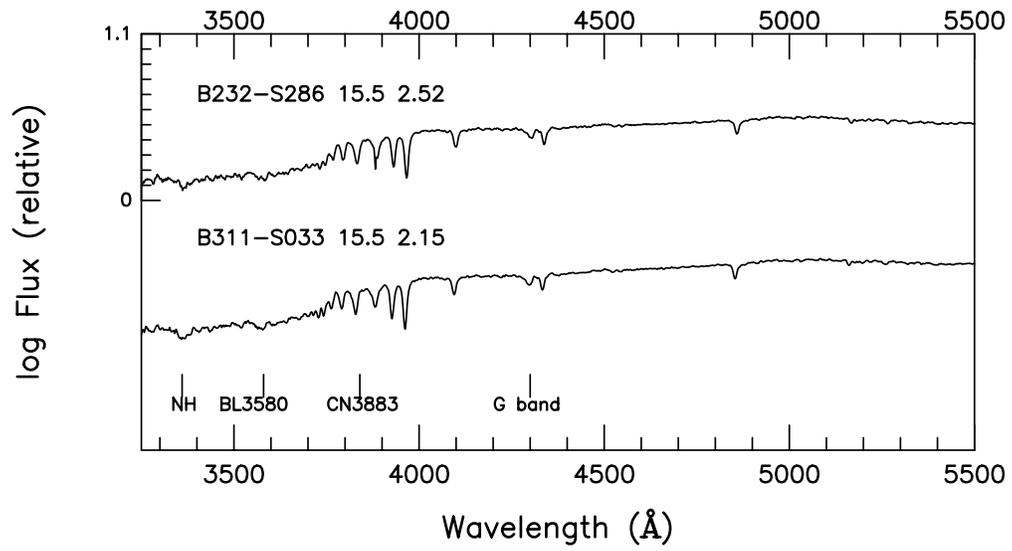}
\figcaption{The spectra for the 2 5-Gyr-old globular clusters found
in our M31 MMT data, plotted on the same log flux scale as in Figure~1.}
\label{fig4}
\end{figure}

\begin{figure}
\epsscale{0.8}
\plotone{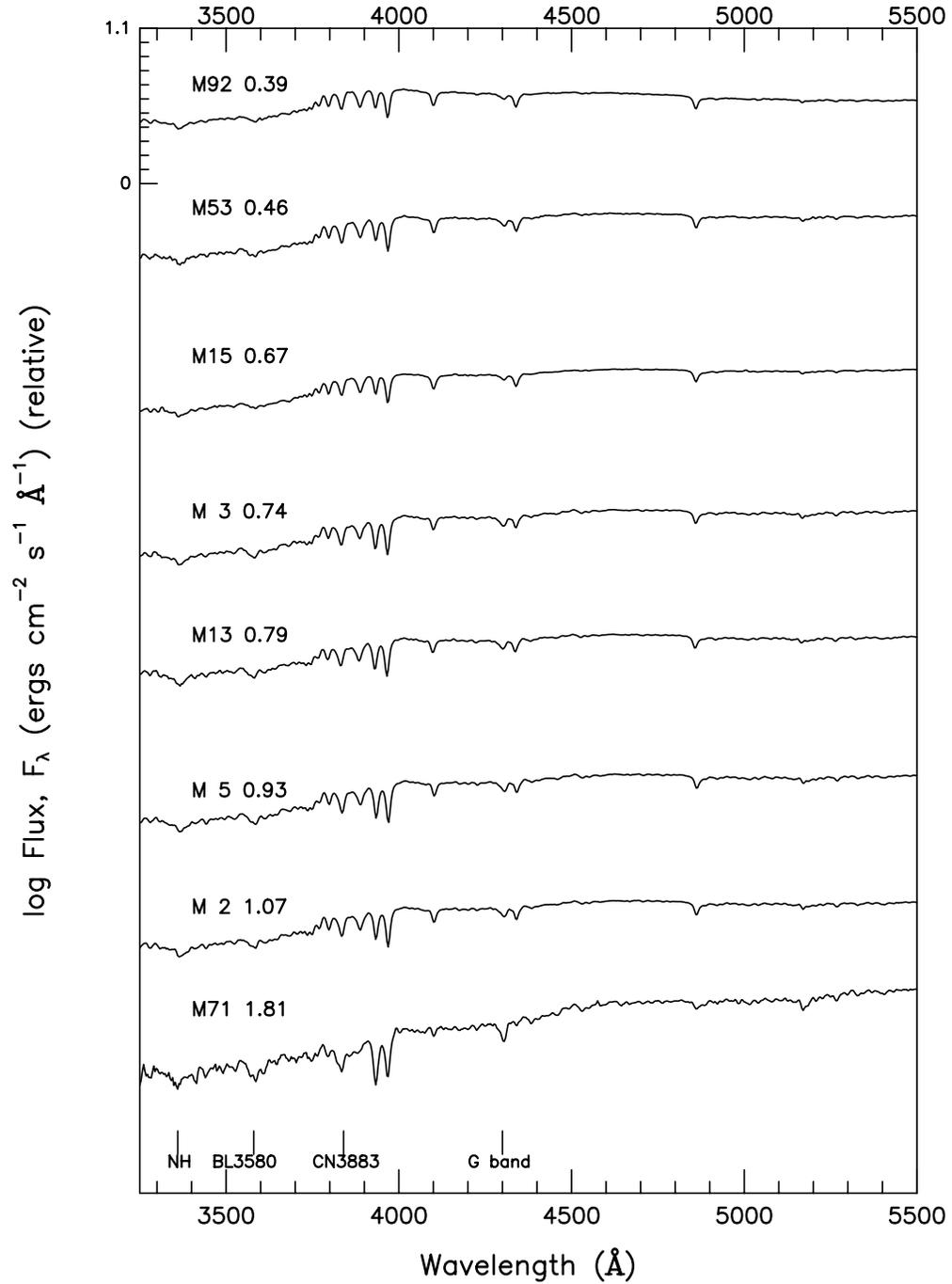}
\figcaption{The spectra for the 8 Milky Way globular clusters from
Li \& Burstein 2003 shown at the same plotting scale as Figs. 1-4 for
comparison. The number by the name of each Milky Way GC is the value
for the Lick index $\rm <Fe>$ for this GC.}
\label{fig5}
\end{figure}

\begin{figure}
\epsscale{0.8}
\plotone{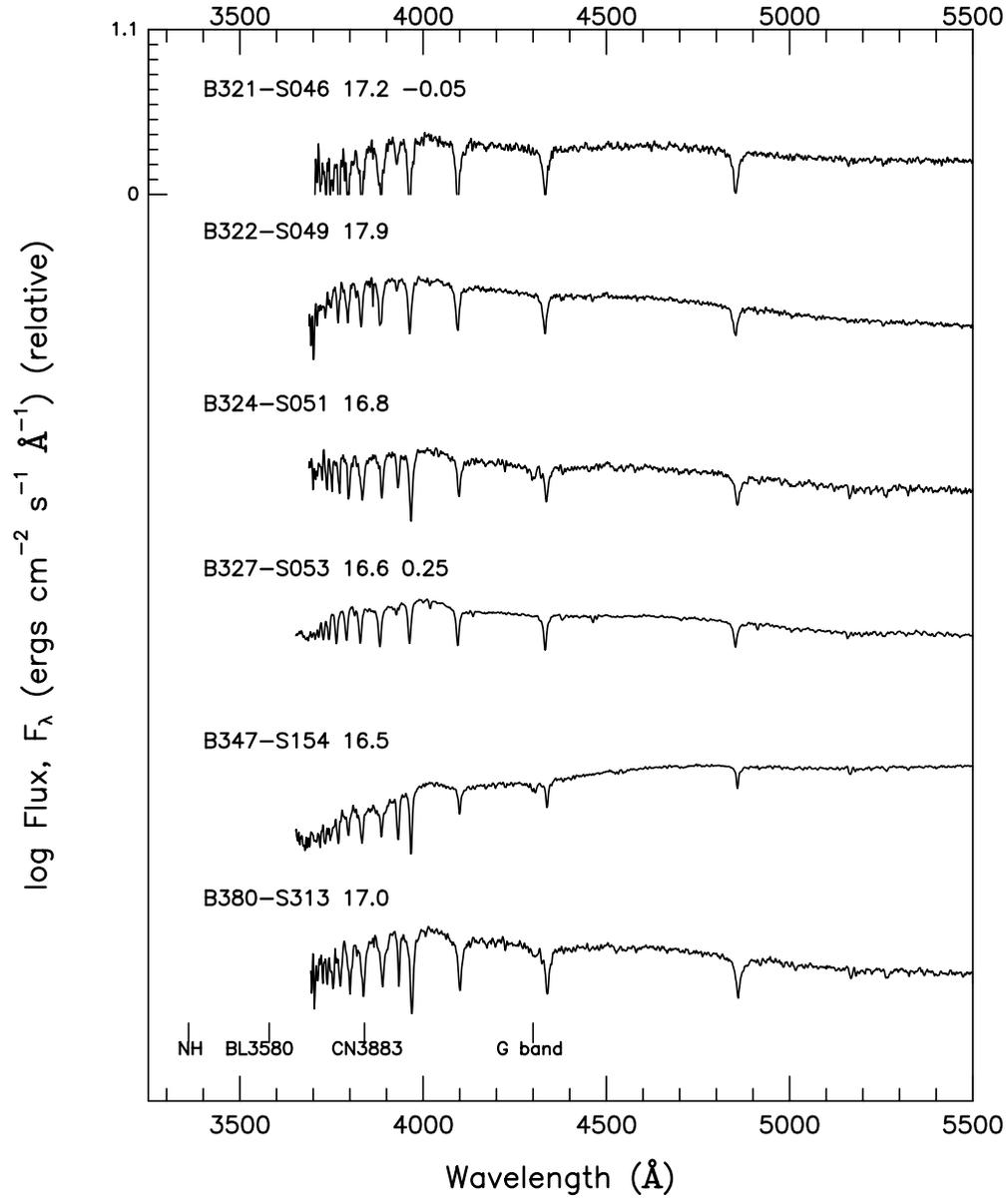}
\figcaption{The Keck spectra of Barmby et al. for six of the
``young?'' M31 GCs that they so indicated in their summary table.
These spectra have been Hanning-smoothed to remove much of the
apparent noise in the original data. Note that these spectra do
not extend much blueward of 3700$\rm \AA$.}
\label{fig6}
\end{figure}

\begin{figure}
\epsscale{0.6}
\plotone{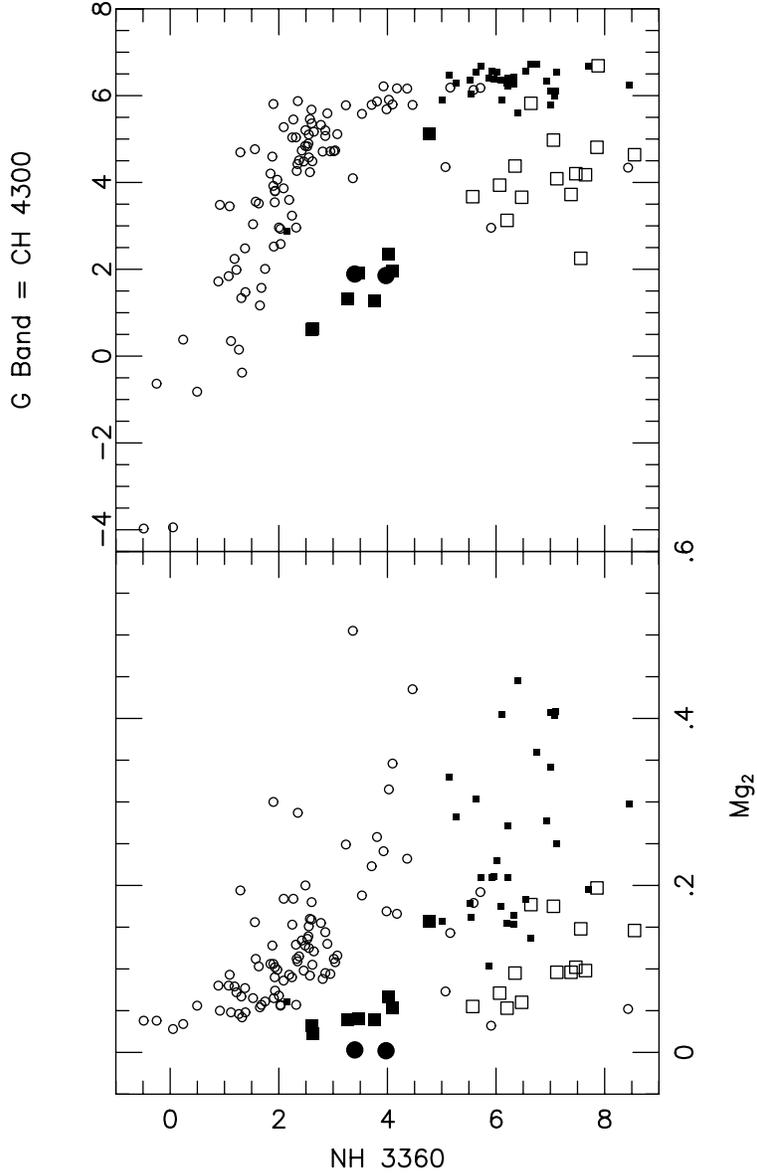}
\figcaption{(top) CH 4300 (the G-band), plotted in units of angstroms,
versus NH, in units of angstroms, for the dwarf stars of Yong Li's
Ph.D. thesis (small open diamonds); the giant stars in his thesis
(small closed circles), and for the Galactic GCs (large closed
squares); for 15 of the older MMT M31 GCs (large open squares;
M31 GC B001-S039 is excluded as its NH measure is too noisy), and for
the two 5 Gyr-old M31 GCs (large closed circles). (bottom) Mg$_2$ 
in units of magnitudes plotted versus NH for the same stars and 
GCs as in the top graph.  All error bars for the GCs are of
comparable sizes to the plotted points.}
\label{fig7}
\end{figure}

\end{document}